\documentclass[12pt]{iopart}
\usepackage{iopams}
\usepackage{graphicx}
\usepackage{amssymb}
\usepackage{amsfonts}
\newcommand{\be}{\begin{equation}}
\newcommand{\ee}{\end{equation}}
\newcommand{\ba}{\begin{eqnarray}}
\newcommand{\ea}{\end{eqnarray}}
\newcommand{\half}{{\textstyle \frac{1}{2}}}
\newcommand{\bp}{{\bar p}}

\newcommand{\bC}{{\hat C}}
\newcommand{\bJ}{{\hat J}}
\newcommand{\bI}{{\mathbf{1}}}
\newcommand{\bU}{{\mathbf{U}}}
\newcommand{\rd}{{\mathrm{d}}}
\newcommand{\re}{{\mathrm{e}}}
\newcommand{\ri}{{\mathrm{i}}}

\begin{document}
\title[Specific Heat of Ising Model with Holes]%
{Specific Heat of Ising Model with Holes: Mathematical Details
Using Dimer Approaches}

\author{Helen Au-Yang and Jacques H.H. Perk}
\address{Department of Physics, Oklahoma State University, 
145 Physical Sciences, Stillwater, OK 74078-3072, USA}
\ead{helenperk@yahoo.com}

\begin{abstract}
In this paper, we use the dimer method to obtain the free energy of Ising models consisting of repeated horizontal strips of width $m$ connected by sequences of vertical strings of length $n$ mutually separated by distance $N$, with $N$ arbitrary, to investigate the effects of connectivity and proximity on the specific heat. The decoration method is used to transform the strings of $n+1$ spins interacting with their nearest neighbors with coupling $J$ into a pair with coupling $\bJ$ between the two spins. The free energy per site is given as a single integral and some results for critical temperatures are derived.
\end{abstract}
\maketitle
\section {Introduction}
\begin{figure}[htb]
  \vspace*{0pt}
  \begin{center}
     \includegraphics[width=0.75\hsize]{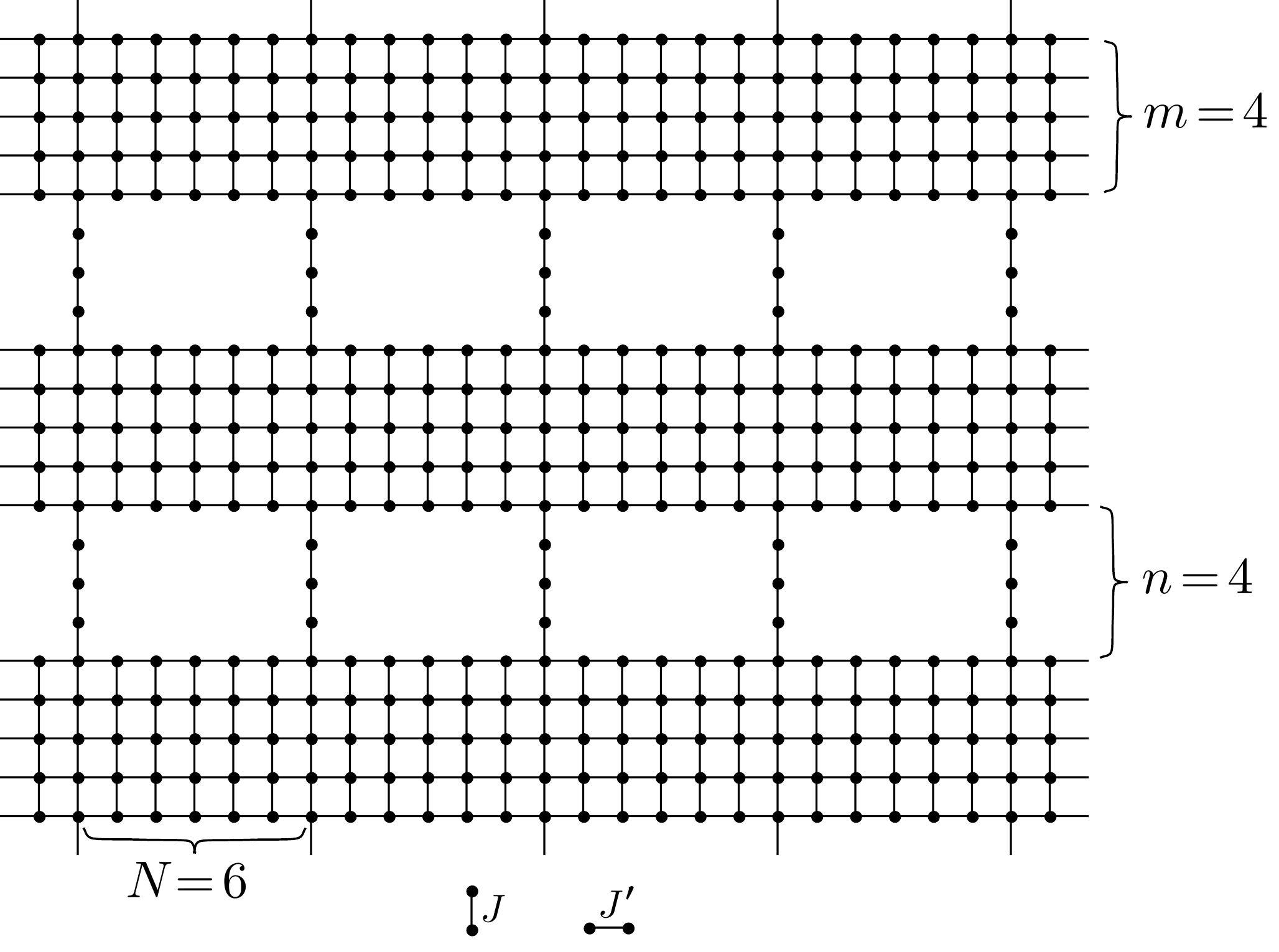} 
  \end{center}
\caption{Part of lattice for case $m=4$, $n=4$ and $N=6$ .}
\label{fig:1}
\end{figure}

We consider an Ising model consisting of $p$ strips of width $m$ and length $N\bar p$, which are connected by $\bar p$ strings of length $n$, which are separated from one another by a distance $N$, see Fig.~\ref{fig:1}, or alternatively Fig.~1 of our previous paper I \cite{HJPih}. We shall consider the ferromagnetic case, for which the free energy per site is independent of boundary conditions in the thermodynamic limit $p,\bar p\to\infty$. Here, unlike paper I, we make the horizontal couplings between the nearest neighbor spins to be $ J'$, different from the vertical couplings $J$, so that one can distinguish the vertical and horizontal correlation lengths. Since this is a two-dimensional Ising model, we expect that the specific heat, which is related to the second derivative of the free energy, to diverge logarithmically at its critical temperature $T_c(N,m,n)$. Various plots have already been given in paper I for the case $J'=J$ omitting the detailed derivations.
In this paper, we will present the mathematical details of the calculation of $T_c(N,m,n)$ and of the free energy as functions of $m,n$ for arbitrary $N$.

\section {Mathematical Details}
The interaction energy $\mathcal{E}$ of the spins $\sigma_{ij}$ is given by
\ba\fl
-\mathcal{E}=\sum_{s=0}^{p-1}\Bigg[\sum_{j=1}^{m+1}{J'}\sum_{k=1}^{\bp N}\sigma_{s(n+m)+j,k}\sigma_{s(n+m)+j,k+1}
+\sum_{j=1}^{m}{J}\sum_{k=1}^{\bp N}\sigma_{s(n+m)+j,k}\sigma_{s(n+m)+j+1,k}
\nonumber\\
+\sum_{j=m+1}^{n+m}{J}\sum_{r=1}^{\bp }\sigma_{s(n+m)+j,rN}\sigma_{s(n+m)+j+1,rN}\Bigg].
\label{int}\ea

\subsection{Decoration Method}
\begin{figure}[htb]
  \vspace*{0pt}
  \begin{center}
     \includegraphics[width=0.75\hsize]{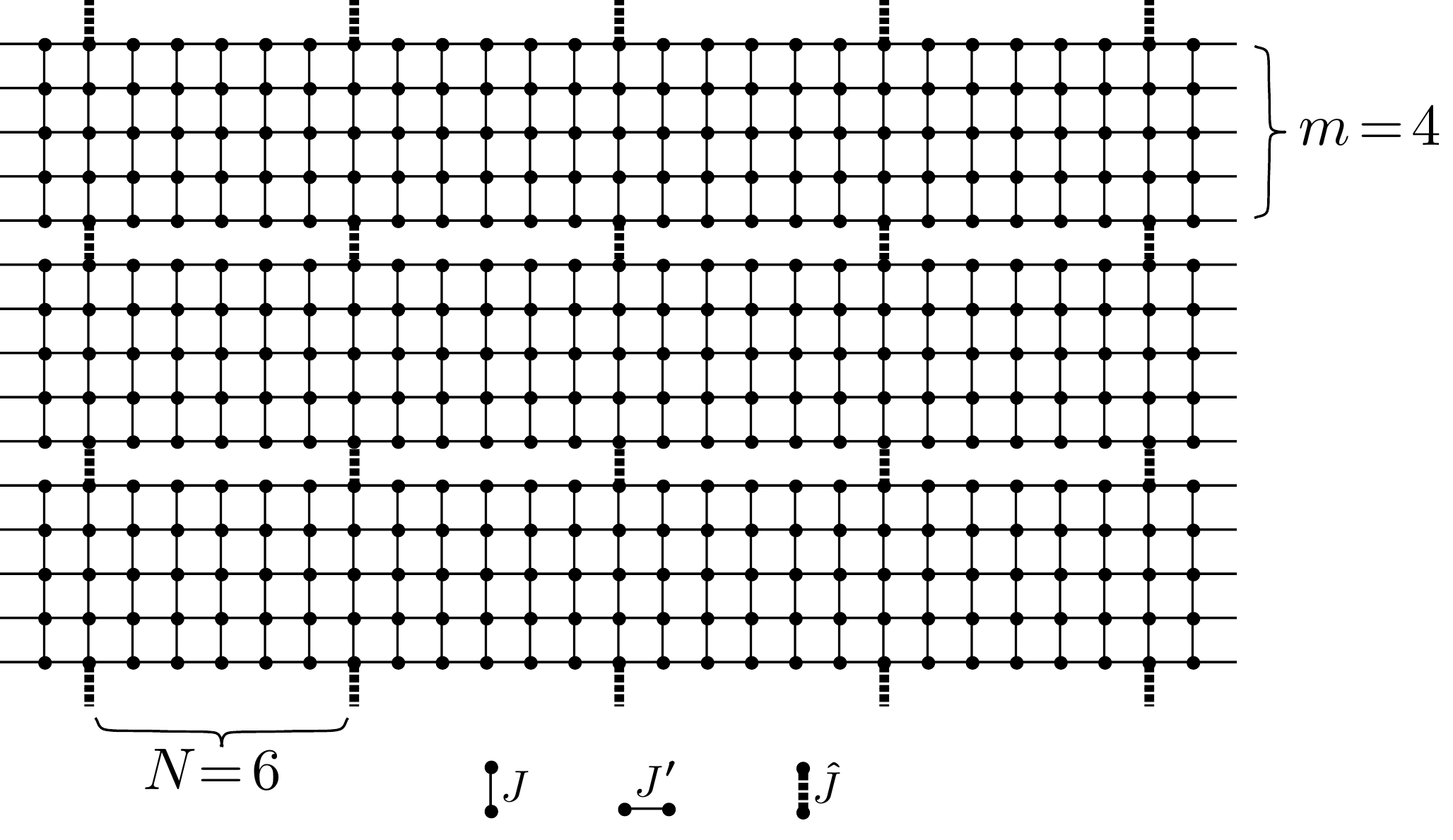} 
    \end{center}
\caption{Decoration method reduction of the lattice in Figure~\ref{fig:1} with all strings
of length $n=4$ replaced by equivalent single bonds of strength $\hat J$.}
\label{fig:2}
\end{figure}

A string of $n+1$ spins $\sigma_1,\ldots,\sigma_{n+1}$, interacting with each other's nearest-neighbor with coupling $J$, can be transformed \cite{Syozi0,Fisher, Syozi} to the spin-pair at the ends of the string with a new coupling $\bJ$, as shown in
equation (73) on page 295 of \cite{Syozi}, or
\be
\mathcal{Z}\equiv\sum_{\sigma_2=\pm 1}\cdots\sum_{\sigma_{n}=\pm 1}\exp\Big[\beta J\sum_{j=1}^{n}\sigma_j\sigma_{j+1}\Big]=\mathcal{I}\exp(\beta{\bJ}\sigma_1\sigma_{n+1}),
\ee
$\beta\equiv1/(k_{\mathrm{B}}T)$.
Going to bond variables $\tau_j=\sigma_j\sigma_{j+1}$ for $j=1,\ldots,n$, $\tau_{n+1}=\sigma_{n+1}$ on the left, and $\hat\tau_1=\sigma_1\sigma_{n+1}$, $\hat\tau_2=\sigma_{n+1}$ on the right,
we find
\ba\fl
\sum_{\sigma_1=\pm 1}\sum_{\sigma_{n+1}=\pm 1}\mathcal{Z}&=
\sum_{\tau_1=\pm 1}\cdots\sum_{\tau_{n+1}=\pm 1}\prod_{j=1}^n\re^{\beta J\tau_j}
=2[2\cosh(\beta J)]^n
\nonumber\\ \fl
&=\mathcal{I}\sum_{\hat\tau_1=\pm 1}\sum_{\hat\tau_2=\pm1}\re^{\beta\hat J\hat\tau_1}
=4\mathcal{I}\cosh(\beta\hat J),
\ea
\ba\fl
\sum_{\sigma_1=\pm 1}\sum_{\sigma_{n+1}=\pm 1}\sigma_1\sigma_{n+1}\mathcal{Z}&=
\sum_{\tau_1=\pm 1}\cdots\sum_{\tau_{n+1}=\pm 1}\prod_{j=1}^n\tau_j\re^{\beta J\tau_j}=
2[(2\sinh(\beta J)]^n
\nonumber\\ \fl
&=\mathcal{I}\sum_{\hat\tau_1=\pm 1}\sum_{\hat\tau_2=\pm1}\hat\tau_1\re^{\beta\hat J\hat\tau_1}
=4\mathcal{I}\sinh(\beta\hat J),
\ea
as all sums trivially factorize upon identifying $\sigma_1\sigma_{n+1}=\prod_{j=1}^n\tau_j$.
Therefore,
\ba
\mathcal{I}=2^{n-1}{[\cosh(\beta J)]^n}/{\cosh(\beta\bJ)},
\label{I}\\
\hat z\equiv\tanh(\beta{\bJ})=z^{n},\quad z\equiv\tanh (\beta J),
\label{tbJ}\ea
which is easily seen to be equivalent to equations (75) and (76) on page 296 in \cite{Syozi}.
Thus each horizontal row of $\bp$ regularly spaced strings of length $n$ becomes a row of $\bp$ regularly spaced vertical bonds $\bJ$ connecting the $p$ strips, as shown in Fig.~\ref{fig:2}. 

More specifically, the model with interaction energy (\ref{int}) is transformed to another model with
interaction energy
\ba\fl
-\mathcal{E}'=\sum_{s=0}^{p-1}\Bigg[\sum_{j=1}^{m+1}{J'}\sum_{k=1}^{\bp N}\sigma_{s(m+1)+j,k}\sigma_{s(m+1)+j,k+1}
\nonumber\\
+\sum_{j=1}^{m}{J}\sum_{k=1}^{\bp N}\sigma_{s(m+1)+j,k}\sigma_{s(m+1)+j+1,k}
+{\hat J}\sum_{r=1}^{\bp }\sigma_{s(m+1),rN}\sigma_{s(m+1)+1,rN}\Bigg].
\label{int2}\ea
Within the $p$ strips the horizontal couplings are still $J'$ and the vertical couplings $J$ for all nearest-neighbor pairs, whereas the $\bp N$ vertical spins pairs on the boundaries of successive strips are not all interacting. Only $\bp$ regularly spaced pairs of them interact with vertical couplings $\bJ$, while also contributing a weight factor $\mathcal{I}$ given by (\ref{I}).

\subsection{Dimer Method}
In order to calculate the free energy per site we shall use here the dimer method of Kasteleyn
and Fisher as presented in chapters 4 and 5 of \cite{MWbk}. For the finite lattice with free boundary
conditions the partition function is expressed in terms of a single Pfaffian, whereas for periodic
boundary conditions a linear combination of four Pfaffians appears.\footnote{The appearance of
four such terms was first encountered by Kaufman using the spinor approach \cite{Kaufman}.}
However, in the thermodynamic limit $p,\bp\to \infty$ the free energy per site is independent of
boundary conditions, so that then the partition function $\mathbf{Z}$ can be replaced by a single Pfaffian of a periodic sparse antisymmetric matrix $\mathbf{A}$ up to some front factors \cite{MWbk}. Here,
\ba
\mathbf{Z}=\mathcal{I}^{p \bp}[2\cosh(\beta J')]^{(m+1)Np \bp} [\cosh(\beta J)]^{mNp\bp}[\cosh(\beta\bJ))]^{p \bp}\,{\mathrm{Pf}}\mathbf{A},
\label{Z}\ea
where $\mathbf{A}$ is of size $4(m+1)Np\bp\times (m+1)Np\bp$. Its non-vanishing elements are firstly the $4\times4$ central matrices associated with each site and given in (2.16) on p.~83 of \cite{MWbk} as
\be
\mathbf{A}(i,j;i,j)=\begin{array}{c}
\begin{array}[t]{ccccc}&\hspace{17pt} R&\hspace{12pt}L&\hspace{5pt}U&\hspace{5pt}D\end{array}\\
\begin{array}{c} R\\L\\U\\D\end{array}\vspace{40pt}
\left(\begin{array}{rrrr}0&1&-1&-1\\
-1&0&1&-1\\
1&-1&0&1\\
1&1&-1&0\end{array}\right)\end{array}=\mathbf{D}.
\label{D}\ee
\vspace*{-30pt} 
The nonzero elements due to the couplings can be deduced from pp.~83--84 in \cite{MWbk}.
The elements related to the horizontal couplings connecting sites $(i,j)$ to $(i,j+1)$ with $1\le i\le (m+1)p$ and $1\le j\le N\bp$ are
\ba
\mathbf{A}(i,j;i,j+1)=-\mathbf{A}(i,j+1;i,j)^T
\nonumber\\
=\begin{array}{c}
\begin{array}[t]{ccccc}&\hspace{12pt} R&\hspace{9pt}L&\hspace{9pt}U&\hspace{5pt}D\end{array}\\
\begin{array}{c} R\\L\\U\\D\end{array}\vspace{40pt}
\left(\begin{array}{cccc}0&\hspace{10pt}{z'}&\hspace{10pt}0&\hspace{10pt}0\\
0&\hspace{10pt}0&\hspace{10pt}0&\hspace{10pt}0\\
0&\hspace{10pt}0&\hspace{10pt}0&\hspace{10pt}0\\
0&\hspace{10pt}0&\hspace{10pt}0&\hspace{10pt}0\end{array}\right)\end{array}=\mathbf{B},\qquad {z'}=\tanh (\beta J'),
\label{B}\ea
\vspace*{-40pt}
\par\noindent
with only a nonzero $RL$ element, as the bond is on the right of the first site and on the left of the second.
The elements related to the vertical bonds connecting sites $(s(m+1)+i,j)$ to $(s(m+1)+i+1,j)$ within the $s$th strip, for $0\le s\le p-1$, $1\le i\le m$, and $1\le j\le N\bp$, are given by
\ba
\fl\mathbf{A}(s(m+1)+i,j;s(m+1)+i+1,j)=-\mathbf{A}(s(m+1)+i+1,j;s(m+1)+i,j)^T
\nonumber\\
=\begin{array}{c}
\begin{array}[t]{ccccc}&\hspace{12pt} R&\hspace{9pt}L&\hspace{5pt}U&\hspace{5pt}D\end{array}\\
\begin{array}{c} R\\L\\U\\D\end{array}\vspace{40pt}
\left(\begin{array}{cccc}0&\hspace{10pt}0&\hspace{10pt}0&\hspace{10pt}0\\
0&\hspace{10pt}0&\hspace{10pt}0&\hspace{10pt}0\\
0&\hspace{10pt}0&\hspace{10pt}0&\hspace{10pt}z\\
0&\hspace{10pt}0&\hspace{10pt}0&\hspace{10pt}0\end{array}\right)\end{array}=\mathbf{C},
\label{C}\ea
\vspace*{-40pt}
\par\noindent
with only a nonzero up-down ($UD$) element, expressing how the bond is connected to the two sites.
Finally, the elements related to the non-vanishing vertical bonds connecting the nearby strips are
\ba
\fl\mathbf{A}(s(m+1),rN;s(m+1)+1,rN)=-\mathbf{A}(s(m+1)+1,rN;s(m+1),rN)^T
\nonumber\\
=\begin{array}{c}
\begin{array}[t]{ccccc}&\hspace{12pt} R&\hspace{9pt}L&\hspace{5pt}U&\hspace{5pt}D\end{array}\\
\begin{array}{c} R\\L\\U\\D\end{array}\vspace{40pt}
\left(\begin{array}{cccc}0&\hspace{10pt}0&\hspace{10pt}0&\hspace{10pt}0\\
0&\hspace{10pt}0&\hspace{10pt}0&\hspace{10pt}0\\
0&\hspace{10pt}0&\hspace{10pt}0&\hspace{10pt}z^n\\
0&\hspace{10pt}0&\hspace{10pt}0&\hspace{10pt}0\end{array}\right)\end{array}
={\mathbf{\bC}}.
\label{bC}
\ea
\vspace*{-40pt}
\par\noindent
for $1\le s\le p$ and $1\le r\le \bp$; this is of the form (\ref{C}) with $z$ replaced by $\hat z=z^n$.

Using (\ref{I}) and $(\mathrm{Pf}\mathbf{A})^2=\det\mathbf{A}$ we can rewrite (\ref{Z}) as
\be
\mathbf{Z}=\left[2^{(m+1)N+n-1}\cosh(\beta J')^{(m+1)N }\cosh(\beta J)^{mN+n}\right]^{p \bp}\,(\det\mathbf{A})^{1/2},
\label{Zz}\ee
thus eliminating the variables $\mathcal{I}$ and $\hat J$, which implicitly depend on $J$ and $\beta=1/k_{\mathrm{B}}T$.
Since we chose the elements of the matrix $\mathbf{A}$ to be doubly periodic with period $m+1$ in the vertical direction and period $N$ in horizontal direction, that is 
\ba
\fl\mathbf{A}(i,j;i',j')=\mathbf{A}(i+m+1,j;i'+m+1,j')=\mathbf{A}(i,j+N;i',j'+N),
 \ea
we can use Fourier transform in both directions.
Consequently, in the thermodynamic limit $p,\bp\to \infty$, we can write the reduced free energy corresponding to partition function (\ref{Zz}) as
\ba
-\beta f&=\lim_{p,\bp\to\infty}\ln\mathbf{Z}/[(m+n)Np\bp]
\nonumber\\
&=f_0(N,m,n) +\frac1{2N(m+n)(2\pi)^2}
\int_0^{2\pi}\rd{\bar\theta}\int_0^{2\pi}\rd\theta\det [\mathbf{U}(\theta,{\bar \theta})],
\label{freeEnergy}\ea
where 
\ba
f_0(N,m,n)&&=\frac{(m+1)N+n-1}{N(m+n)}\ln 2+\frac{m+1}{(m+n)}\ln\cosh(\beta J')
\nonumber\\
&&+\frac{n+mN}{N(m+n)}\ln\cosh(\beta J),
\label{f0}\ea
and $\mathbf{U}(\theta,{\bar \theta})$ is a $4(m+1)N\times4(m+1)N$ matrix given by
\begin{figure}[tb]
  \vspace*{0pt}
  \begin{center}
     \includegraphics[width=0.35\hsize]{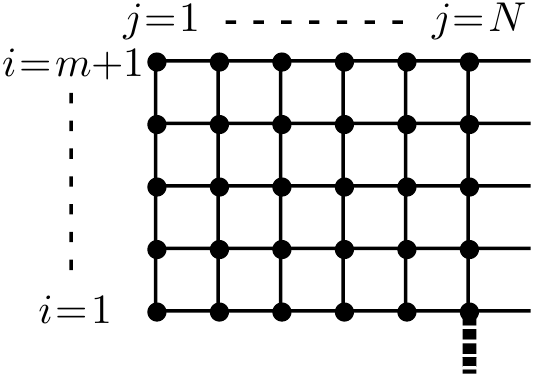} 
    \end{center}
\caption{Choice of unit cell of the lattice in Figure~\ref{fig:2} used in the construction
of the Pfaffian in the text. }
\label{fig:3}
\end{figure}
\be
\mathbf{U}(\theta,{\bar \theta})=\left(\begin{array}{cccccc}
X&Y&0&\cdots&0&-Z^T\re^{-\ri\theta}\\
-Y^T&X&Y&\cdots&0&0\\
0&-Y^T&X&Y&&0\\
\vdots&\vdots&\ddots&\ddots&\ddots&\vdots\\
0&0&\cdots&-Y^T&X&Y\\
Z\re^{\ri\theta}&0&\cdots&0&-Y^T&X
\end{array}\right)_{\!\!\!(m+1)\times(m+1)}
\label{U}\ee
with elements $\mathbf{U}(\theta,{\bar \theta})_{i,i'}$ expressed in terms of the
$4N\times4N$ matrices $X,Y,Z$ and the zero matrix 0. From Fig.~\ref{fig:3} we see
that the only nonzero elements are the ones with $i'=i,i\pm1$ mod $m+1$, whereas
$\mathbf{U}(\theta,{\bar \theta})_{1,m+1}$ and $\mathbf{U}(\theta,{\bar \theta})_{m+1,1}$ pick up a factor $\re^{\pm\ri\theta}$ from the Fourier transform,
as the corresponding interactions connect neighboring unit cells.\footnote{Compare also equations (2.10) and (2.11) on p.~1240 of \cite{HAYFisherFerd}.}
Because the structure of each row in Fig.~\ref{fig:3} is the same, we see that all diagonal elements $\mathbf{U}(\theta,{\bar \theta})_{i,i}$
have to be the same matrix
\ba
X=\left(\begin{array}{ccccc}
\mathbf{D}&\mathbf{B}&0&\cdots&-\mathbf{B}^T\re^{-\ri{\bar\theta}}\\
-\mathbf{B}^T&\mathbf{D}&\mathbf{B}&\cdots&0\\
\vdots&\ddots&\ddots&\ddots&\vdots\\
0&\cdots&-\mathbf{B}^T&\mathbf{D}&\mathbf{B}\\
\mathbf{B}\re^{\ri{\bar\theta}}&0&\cdots&-\mathbf{B}^T&\mathbf{D}\end{array}\right),
\label{X}\ea
with elements $X_{j,j'}$ that are $4\times4$ matrices, nonzero only for $j'=j,j\pm1$ mod $N$.
Now $X_{jj}=\mathbf{D}$, $X_{j,j+1}=\mathbf{B}$ and $X_{1N}$ and $X_{N1}$ pick up
factors $\re^{\pm\ri\bar\theta}$ from the Fourier transform.
The off-diagonal elements of (\ref{U}) can only involve vertical interactions, so their nonvanishing elements must have $j'=j$, leading to the diagonal forms
\ba
Y=\left(\begin{array}{cccc}
\mathbf{C}&\hspace{10pt}0&\cdots&\hspace{10pt}0\\
0&\hspace{10pt}\mathbf{C}&\cdots&\hspace{10pt}0\\
\vdots&\hspace{10pt}\ddots&\ddots&\hspace{10pt}\vdots\\
0&\hspace{10pt}\cdots&0&\hspace{10pt}\mathbf{C}\end{array}\right),
\quad
Z=\left(\begin{array}{cccc}
0&\hspace{10pt}0&\cdots&\hspace{10pt}0\\
0&\hspace{10pt}0&\cdots&\hspace{10pt}0\\
\vdots&\hspace{10pt}\ddots&\ddots&\hspace{10pt}\vdots\\
0&\hspace{10pt}\cdots&0&\hspace{10pt}\mathbf{\bC}\end{array}\right),
\label{YZ}\ea
with $\mathbf{C}$ and $\mathbf{\bC}$ the $4\times 4$ matrices given in (\ref{C}) and (\ref{bC}). 

Now the matrix $\mathbf{U}(\theta,{\bar \theta})$ has been fully defined in (\ref{U})--(\ref{YZ}), the
next problem is how to calculate its determinant. We can see that the matrices at the up-right and down-left corners of (\ref{U}) complicate the calculation of this determinant. There are two known
ways of how to proceed. We have first used the iteration method as given on pages 118--121
in \cite{MWbk} and also used in \cite{HAYBM,Hamm} to calculate the determinant directly.
Since this calculation was rather messy, and there was a great deal of cancellations, we have
also used another method for comparison and checking, namely the one described in \cite{HAYFisherFerd}. In that method, instead of calculating the determinant of
matrix $\mathbf{U}$ itself, we have to take out the matrix $\mathbf{U}_0$ of the perfect uniform Ising lattice, for which the matrices $Z$ and $Z^T$ in $\mathbf{U}$ on the corners of (\ref{U}) are replaced by $Y$ and $Y^T$. The resulting difference $\mathbf{U}-\mathbf{U}_0$ is a $2N\times 2N$ matrix, so that we only need to calculate the determinant due to this $2N\times 2N$ matrix. Again there is a great deal of cancellations, yielding the same final result with much less work.

In the next section 3 we shall only outline the first method, without presenting the details, which are
extremely lengthy and tedious already for $N=3$. The second method shall be explained in detail in section 4.

\section{Iteration Method}
Since $X$ in (\ref{X}) is nearly cyclic, it can be block-diagonalized by Fourier transform as
\be
 (R^{-1}\otimes \bI_4)X(R\otimes \bI_4)=\left(\begin{array}{cccc}
{\hat\mathbf{D}}(\phi_0)&\hspace{10pt}0&\cdots&\hspace{10pt}0\\
0&\hspace{10pt}{\hat\mathbf{D}}(\phi_1)&\cdots&\hspace{10pt}0\\
\vdots&\hspace{10pt}\ddots&\ddots&\hspace{10pt}\vdots\\
0&\hspace{10pt}\cdots&0&\hspace{10pt}{\hat\mathbf{D}}(\phi_{N-1})\end{array}\right),
\label{dX}\ee
where
\be
{\hat\mathbf{D}}(\phi_k)=\mathbf{D}+\re^{\ri\phi_k}\mathbf{B}-\re^{-\ri\phi_k}\mathbf{B}^T,
\label{hD}\ee
and the $N\times N$ matrix $R$ is given by
\ba
R_{jk}=\frac{\re^{\ri j\phi_{k}}}{\sqrt{N}},\quad (R^{-1})_{k j}=\frac{\re^{-\ri j\phi_{k}}}{\sqrt{N}},
\quad \phi_k\equiv\frac{2\pi k+{\bar \theta}}N,\nonumber\\
j=1,\cdots,N,\quad k=0,\ldots,N-1.
\label{Fourier}\ea
Since $Y=\textbf{1}_N\otimes\mathbf{C}$, we have
\be (R^{-1}\otimes \bI_4)Y(R\otimes \bI_4)=Y.
\ee
But 
\be
 (R^{-1}\otimes \bI_4)Z(R\otimes \bI_4)={\hat Z}\otimes{\mathbf{\bC}}/N,
\label{hCZ}\ee
in which $ {\hat Z}$ is an $N\times N$ matrix, with all its elements being equal to 1, that is,
\be
{\hat Z}={\left(\begin{array}{cccc}
1&1&\hspace{10pt}\cdots\hspace{10pt}&1\\
1&1&\hspace{10pt}\cdots\hspace{10pt}&1\\
\vdots&\vdots&\ddots&\vdots\\
1&1&\cdots&1\end{array}\right).}_{\hspace{-0.8em}N\times N}
\label{hZ}\ee
Using (\ref{D}) and (\ref{B}), we rewrite (\ref{hD}) as
\be
{\hat\mathbf{D}}(\phi_k)=\begin{array}{c}
\begin{array}[t]{ccccc}&\hspace{18pt} R&\hspace{7pt}L&\hspace{5pt}U&\hspace{5pt}D\end{array}\\
\begin{array}{c} R\\L\\U\\D\end{array}\vspace{40pt}
\left(\begin{array}{rrrr}0&a_k&-1&-1\\
-{\bar a_k}\!\!\!&0&1&-1\\
1&-1&0&1\\
1&1&-1&0\end{array}\right)\end{array},\quad 
\begin{array}{c}a_k=1+\re^{\ri\phi_k}{z'},\\ {\bar a_k}=1+\re^{-\ri\phi_k}{z'},\end{array}
\label{hD1}\ee
compare equation (3.3$a$) on page 118 of \cite{MWbk}. Looking back at (\ref{C}) and (\ref{bC})
we see that these matrices and their transposes, have zero elements both in their UD rows and their RL columns. Therefore, to make $\mathbf{U}(\theta,{\bar \theta})$ block-triangular, we follow the method on page 119 of \cite{MWbk} and eliminate the UD-RL elements of each ${\hat D}(\phi_k)$ in (\ref{dX})--- and given in (\ref{hD1})---by multiplying from the left by the matrix 
\be \mathbf{H}=\left(\begin{array}{cccc}H(\phi_0)&0&\cdots&0\\
0&H(\phi_1)&\cdots&0\\
\vdots&\ddots&\ddots&\vdots\\
0&\cdots&0&H(\phi_{N-1})\end{array}\right),
\ee
with
\be
H(\phi_k)=\left(\begin{array}{ccrr}
1&0&0&0\\
0&1&0&0\\
a_k^{-1}&{\bar a_k}^{-1}&1&0\\
-a_k^{-1}\!\!&{\bar a_k}^{-1}&0&1\end{array}\right),\quad \det[H(\phi_k)]=1,
\ee
It can be easily verified that
\be
\fl H(\phi_k)\hat D(\phi_k)=\left(\begin{array}{ccrr}
0&a_k&-1&-1\\
-{\bar a_k}&0&1&-1\\
0&0&u_k&-v_k\\
0&0&v_k&-u_k\end{array}\right),\quad
\begin{array}{c}u_k={\bar a_k}^{-1}-a_k^{-1}
=\displaystyle{\frac{2\ri{z'}\sin\phi_k}{|1+{z'}\re^{\ri\phi_k}|^2}},\vspace{10pt}\\
v_k={\bar a_k}^{-1}+a_k^{-1}-1=
\displaystyle{\frac{1-{z'}^2}{|1+{z'}\re^{\ri\phi_k}|^2},}\end{array}
\label{HhD}\ee
compare (3.4$a$) in \cite{MWbk},
and
\be
H(\phi_k)\,\mathbf{C}=\mathbf{C},\quad H(\phi_k)\,\mathbf{\hat C}=\mathbf{\hat C},
\label{ChC}\ee
so that
\be
\mathbf{H}Y=Y,\quad\mathbf{H}{\hat Z}\otimes{\mathbf{\bC}}/N={\hat Z}\otimes{\mathbf{\bC}}/N.
\label{YhZ}\ee
Consequently, 
\ba
\fl\det[\mathbf{U}(\theta,{\bar \theta})]=
\det[(\bI_{m+1}\otimes R^{-1}\otimes \bI_4)\mathbf{U}(\theta,{\bar \theta})
(\bI_{m+1}\otimes R\otimes \bI_4]
\label{U1}\\
=\det[(\bI_{m+1}\otimes\mathbf{H})(\bI_{m+1}\otimes R^{-1}\otimes\bI_4)
\mathbf{U}(\theta,{\bar \theta})(\bI_{m+1}\otimes R\otimes \bI_4]
\label{U2}\\
=\prod_{k=0}^{N-1}|1+{z'}\re^{\ri\phi_k}|^{2(m+1)}\det[\mathbf{V}(\theta,{\bar\theta})].
\label {U3}\ea
In (\ref{U1}), the matrices $X$ in (\ref{U}) are transformed into block-diagonal form (\ref{dX}), and then, by multiplying by the determinant-1 matrix $\bI_{m+1}\otimes\mathbf{H}$ as shown in (\ref{U2}), 
this $X$  is further transformed into block-triangular form with only $2\times2$ diagonal blocks. From (\ref{ChC}) and (\ref{YhZ}) one sees that the matrix $Y$ is invariant under these transforms, whereas $Z$ becomes more complicated, yet preserving the property that all elements in all RL rows and columns are identically zero. Thus, for the calculation of its determinant, the matrix $\mathbf{U}$ becomes triangular with two diagonal blocks of size $2N(m+1)\times2N(m+1)$. The RL block only comes from the matrix $X$ and its determinant is easy to calculate with the result being the product in (\ref{U3}). Hence, the calculation of the determinant of the $4N(m+1)\times4N(m+1)$ matrix $\mathbf{U}$ is reduced to the calculation of the determinant of the other $2N(m+1)\times2N(m+1)$ UD block, which we called $\mathbf{V}(\theta,{\bar\theta})$ in (\ref{U3}).

This $\mathbf{V}(\theta,{\bar\theta})$ can be written as a tridiagonal matrix similar to (3.9) on page 120 of \cite{MWbk}, plus $2N^2$ elements coming from ${\hat Z}$ in (\ref{hZ}). These extra elements made the further calculation of the determinant of $\mathbf{U}$ very tedious. Nevertheless we were able to calculate the sparse determinant even for $N=3$, in which case the derivation is particularly lengthy and messy. Rather than giving further details, we shall next describe in detail the second method that led to the same final results for $N=1,2,3$, but after a simple modification also gives the answer for general $N$.

\section{Method of Subtraction}
As an other application of  the method described in \cite{HAYFisherFerd}, we shall now calculate the determinant of the matrix $\mathbf{U}$ by taking out matrix $\mathbf{U}_0$ of the perfect uniform Ising lattice.\footnote{This had been done earlier in \cite{MPW} in order to calculate correlation functions by the dimer method.}
That is, we replace $Z$ in (\ref{U}) by $Y$ and find
\be
\bU_0(\theta,{\bar \theta})={\left(\begin{array}{cccccc}
X&Y&0&\cdots&0&-Y^T\re^{-\ri\theta}\\
-Y^T&X&Y&\cdots&0&0\\
0&-Y^T&X&Y&&0\\
\vdots&\vdots&\ddots&\ddots&\ddots&\vdots\\
0&0&\cdots&-Y^T&X&Y\\
Y\re^{\ri\theta}&0&\cdots&0&-Y^T&X
\end{array}\right),}_{\hspace{-0.8em}(m+1)\times(m+1)}
\label{U0}\ee
compare (2.21) in \cite{HAYFisherFerd}.
Since $\bU_0(\theta,{\bar \theta})$ is nearly cyclic, its determinant is easily calculable. It becomes block-diagonal with $4\times4$ blocks after Fourier transform (\ref{Fourier}) and a similar one with
$N$ replaced by $m+1$ and $\bar\theta$ by $\theta$. The result for the determinant is given in (2.17)--(2.19) of \cite{HAYFisherFerd}, namely\footnote{In order to compare with \cite{HAYFisherFerd,MPW}, identify $z_1=z'$, $z_2=z$, $\phi_1=\phi$, $\phi_2=\psi$, $\theta_1=\bar\theta$, $\theta_2=\theta$.}
\be
\ln|\bU_0(\theta,{\bar \theta})|=\sum_{\phi}\sum_{\psi}\ln\Delta(\phi, \psi),
\label{sU0}\ee
where 
\ba
\Delta(\phi, \psi)=(1+z^2)(1+z'{}^2)-2z'(1-z^2)\cos\phi-2z(1-z'{}^2)\cos\psi
\nonumber\\
\hspace{42pt}=z(1-z'{}^2)[\alpha(\phi)+\alpha(\phi)^{-1}-2\cos\psi]
\label{Delta}\ea
with $\alpha(\phi)$ the roots of the implied second degree equation\footnote{This $\alpha(\phi)$ is real, as the absolute minimum of the right-hand side of (\ref{alpha}) is 2 for $0\le z,z'\le1$.}
\be 
\alpha(\phi)+\alpha(\phi)^{-1}=[(1+z^2)(1+z'{}^2)-2z'(1-z^2)\cos\phi]\Big/[z(1-z'{}^2)];
\label{alpha}
\ee
the sums in (\ref{sU0}) are over all the values of $\phi$ and $\psi$ given by
\ba
\phi=(2\pi k+\bar\theta)/N,\quad k=0,\cdots,N-1,
\nonumber\\
\psi=(2\pi j+\theta)/(m+1),\quad j=0,\cdots,m.
\label{psi}\ea
It is well-known that by replacing the sums in (\ref{sU0}) by integrals one recovers Onsager's free energy result. In order to get that result in the form we use later, we first carry out the integral over $\theta$ in (\ref{freeEnergy}) giving
\ba
\fl\int_0^{2\pi}\rd\theta\ln|\bU_0(\theta,{\bar \theta})|=\sum_{\phi}\int_0^{2\pi}\rd\theta\sum_{\psi}\ln\Delta(\phi, \psi)
\nonumber\\
\fl=(m+1)\sum_{\phi}\sum_{j=0}^m\int_{2\pi j/(m+1)}^{2\pi(j+1)/(m+1)}\rd\psi\ln\Delta(\phi, \psi)=(m+1)\sum_{\phi}\int_0^{2\pi}\rd\psi\ln\Delta(\phi, \psi).
\ea
Then, using the formula
\be
\frac 1{2\pi}\int_0^{2\pi}\rd x\ln(a-b\cos x)=\ln\Big[\half\Big(a+\sqrt{a^2-b^2}\Big)\Big],
\label{integral}\ee
and (\ref{Delta}), we find
\ba
\int_0^{2\pi}\!\!\rd\theta\ln|\bU_0(\theta,{\bar \theta})|=2\pi(m+1)\sum_{\phi}\left[\ln\left(z'(1-z^2)\right)+\ln\alpha(\phi)\right],
\ea
so that
\ba\fl
\int_0^{2\pi}\!\!\rd\bar\theta\int_0^{2\pi}\!\!\rd\theta\ln\!|\bU_0(\theta,{\bar \theta})|=
(2\pi)^2(m+1)N\ln[z'(1-z^2)]+2\pi\!\int_0^{2\pi}\!\!\!\rd\bar\theta\sum_{\phi}\ln\alpha(\phi)^{m+1}.
\label{intU0}\ea

Now, as in (2.21) of \cite{HAYFisherFerd}, we write
\ba\fl
|\bU(\theta,{\bar \theta})|=|\bU_0(\theta,{\bar \theta})|\cdot|\bI+\bU^{-1}_0(\bU-\bU_0)|
=|\bU_0(\theta,{\bar \theta})|\cdot|\bI'+\mathbf{G}(\theta,{\bar \theta})\mathbf{y}|,
\label{dU}\ea
where $\mathbf{y}$ denotes the square submatrix consisting of the non-vanishing rows and columns of the difference $\Delta \bU=\mathbf{U}-\mathbf{U}_0$; and $\bI'$ and $\mathbf{G}$ are the corresponding submatrices of $\bI$ and $\bU^{-1}_0$, as the other nonzero elements of $\bU^{-1}_0\Delta U$ do not contribute to the determinant.
From (\ref{U0}) and (\ref{U}), we find that the difference $\Delta \bU=\mathbf{U}-\mathbf{U}_0$ is non-zero only at the lower-left and upper-right corners in these equations, namely
\ba
(Z-Y)\re^{\ri\theta}&=&\re^{\ri\theta}{\left(\begin{array}{cccc}
-C&\hspace{10pt}0&\cdots&\hspace{10pt}0\\
0&\hspace{10pt}-C&\cdots&\hspace{10pt}0\\
\vdots&\hspace{10pt}\ddots&\ddots&\hspace{10pt}\vdots\\
0&\hspace{10pt}\cdots&0&\hspace{10pt}{\bC}-C\end{array}\right),}_{\hspace{-0.8em}4N\times4N}
\nonumber\\
(Y^T-Z^T)\re^{-\ri\theta}&=&\re^{-\ri\theta}{\left(\begin{array}{cccc}
C^T&\hspace{10pt}0&\cdots&\hspace{10pt}0\\
0&\hspace{10pt}C^T&\cdots&\hspace{10pt}0\\
\vdots&\hspace{10pt}\ddots&\ddots&\hspace{10pt}\vdots\\
0&\hspace{10pt}\cdots&0&\hspace{10pt}C^T-{\bC}^T\end{array}\right).}_{\hspace{-0.8em}4N\times4N}
\ea
From (\ref{C}) and (\ref{bC}), we see that these matrices are non-zero only between the U D rows and columns, thus
the non-zero submatrix of $\Delta \bU$ is $2N\times 2N$ matrix given by
\ba
 \fl{\mathbf{y}}=
 \nonumber\\
 \vspace{-25pt}
\fl\!\!\!\begin{array}{c}
\begin{array}[t]{ccccccccc}
&\hspace{45pt} 1D&\hspace{12pt}2D&\hspace{5pt}\cdots&\hspace{15pt}ND
&\!\!\!(mN\!+\!1)U&\!\!\!(mN\!+\!2)U&\!\cdots\!&(m\!+\!1)NU\end{array}\\
\!\!\!\!\begin{array}{c} 1D\\2D\\\vdots\\ND\\(mN\!+\!1)U\\(mN\!+\!2)U\\\vdots\\(m\!+\!1)NU\end{array}
\left(\begin{array}{cccccccc}0&0&\cdots&0&\re^{-\ri\theta}z&\hspace{20pt}0&\hspace{5pt}\cdots&0\\
0&0&\cdots&0&0&\hspace{10pt}\re^{-\ri\theta}z&\hspace{5pt}\cdots&0\\
\vdots&\vdots&\ddots&\vdots&\vdots&\hspace{15pt}\vdots&\hspace{5pt}\ddots&\vdots\\
0&0&\cdots&0&0&\hspace{15pt}0&\hspace{5pt}\cdots&\re^{-\ri\theta}(z-z^n)\\
-\re^{\ri\theta}z&0&\cdots&0&0&\hspace{15pt}0&\hspace{5pt}\cdots&0\\
0&-\re^{\ri\theta}z&\cdots&0&0&\hspace{15pt}0&\hspace{5pt}\cdots&0\\
\vdots&\vdots&\ddots&\vdots&\vdots&\hspace{15pt}\vdots&\hspace{5pt}\ddots&\vdots\\
0&0&\cdots&\re^{\ri\theta}(z^n-z)&0&\hspace{15pt}0&\hspace{5pt}\cdots&0\\
\end{array}\!\!\!\right)\end{array}\hspace{-5pt}.
\label{y}\ea
Indeed, numbering the vertices $(i,j)$ in figure \ref{fig:3} by $(i-1)N+j$ and noting that the vertices in the top row are connected by their U ports with the corresponding ones in the bottom row by their D ports, we receive (\ref{y}), or equivalently
\ba
{\mathbf{y}}_{\ell,D;\ell',U}=(z-z^n\delta_{\ell,N})\re^{-\ri\theta}\delta_{\ell,\ell'},\nonumber\\
{\mathbf{y}}_{\ell,U;\ell',D}=(z^n\delta_{\ell',N}-z)\re^{\ri\theta}\delta_{\ell,\ell'},\nonumber\\
{\mathbf{y}}_{\ell,U;\ell',U}={\mathbf{y}}_{\ell,D;\ell',D}=0,
\label{yy}\ea
with $\ell,\ell'=1,\ldots,N$.

The matrix $\mathbf{G}$ is the submatrix of $\bU^{-1}_0$ with same rows and columns as that of ${\mathbf{y}}$. The inverse of the nearly cyclic matrix $\bU_0$ is well known and is given in \cite{HAYFisherFerd,MPW}. The double Fourier transform of $\bU^{-1}_0$ is diagonal with its diagonal elements the $4\times4$ matrices (10) in \cite{MPW}, with their inverses given in (26) of \cite{MPW}, so that $\bU^{-1}_0$ is then the inverse Fourier transform (25) in \cite{MPW}.
More specifically, we find from equations (2.23) to (2.25) in \cite{HAYFisherFerd} that 
\ba
\fl{\mathbf{G}}_{\ell,D;\ell',D}=\bU_0^{-1}(\theta,{\bar \theta})_{\ell,D;\ell',D}=\frac 1{N(m+1)}\sum_{\phi}\sum_{\psi}\re^{\ri(\ell-\ell')\phi}\left[\frac{2\ri z'\sin\phi}{\Delta(\phi, \psi)}\right],
\label{Gdd}\\
\fl{\mathbf{G}}_{\ell,U;\ell',U}=\bU_0^{-1}(\theta,{\bar \theta})_{\ell+mN,U;\ell'+mN,U}
=-{\mathbf{G}}_{\ell,D;\ell',D},
\label{Guu}\\
\fl{\mathbf{G}}_{\ell,D;\ell',U}=\bU_0^{-1}(\theta,{\bar \theta})_{\ell,D;\ell'+mN,U}
\nonumber\\
\hspace{-27pt}=\frac 1{N(m+1)}\sum_{\phi}\sum_{\psi}\re^{\ri(\ell-\ell')\phi-\ri m\psi} 
\left[\frac{z\re^{-\ri\psi}|1+z'\re^{\ri\phi}|^2-(1-z'{}^2)}{\Delta(\phi, \psi)}\right],
\label{Gdu}\\
\fl{\mathbf{G}}_{\ell,U;\ell',D}=\bU_0^{-1}(\theta,{\bar \theta})_{\ell+mN,U;\ell',D}=-\overline{\mathbf{G}}_{\ell',D;\ell,U},
\label{Gud}\ea
with $\ell$ and $\ell'$ interchanged in (\ref{Gud}) and where $\bar x$ denotes the complex conjugate of $x$.

One of the sums in (\ref{Gdd}) and (\ref{Gdu}) can be carried out. As $m$ is arbitrary, and $N$ small, we choose to carry out the sum over $\psi$. From (\ref{Delta}), we can write
\ba
\Delta(\phi, \psi)=z(1-z'{}^2)\re^{-\ri\psi}[1-\re^{\ri\psi}\alpha(\phi)][\re^{\ri\psi}\alpha(\phi)^{-1}-1],
\label{delta1}\ea
so that
\ba
\fl\frac 1{\Delta(\phi, \psi)}\!=\!\frac 1{z(1-z'{}^2)[\alpha(\phi)-\alpha(\phi)^{-1}]}
\left[\frac 1{1-\alpha(\phi)^{-1}\re^{\ri\psi}}-\frac 1{1-\alpha(\phi)\re^{\ri\psi}}\right],
\nonumber\\
\fl\frac{\re^{-\ri\psi}}{\Delta(\phi, \psi)}\!=\!\frac 1{z(1-z'{}^2)[\alpha(\phi)-\alpha(\phi)^{-1}]}\left[\frac {\alpha(\phi)^{-1}}{1-\alpha(\phi)^{-1}\re^{\ri\psi}}-\frac {\alpha(\phi)}{1-\alpha(\phi)\re^{\ri\psi}}\right].
\label{delta2}\ea
Using the second member of (\ref{psi}), we find
\be
\fl\frac 1{m+1}\sum_{\psi}\frac 1{1-x\re^{\ri\psi}}=\frac 1{m+1}\sum_{\psi}
\sum_{j=0}^\infty(x\re^{\ri\psi})^j=\sum_{k=0}^\infty(x^{m+1}\re^{\ri\theta})^k
=\frac1{1-x^{m+1}\re^{\ri\theta}},
\label{sum}\ee
as only the terms with $j=k(m+1)$ survive.

Consequently, from (\ref{delta2}), and using (\ref{sum}), we obtain
\ba
S_0\equiv\frac 1{m+1}\sum_{\psi}\frac 1{\Delta(\phi, \psi)}=\frac 1{z(1-z'{}^2)}
\frac{\mathcal{H}_{m+1}(\phi)}{\mathcal{R}(\phi,\theta)},
\label{s0}\\
S_{-1}\equiv{\bar S}_{1}\equiv\frac 1{m+1}\sum_{\psi}\frac {\re^{-\ri\psi}}{\Delta(\phi, \psi)}
=\frac 1{z(1-z'{}^2)}\frac{\re^{-\ri\theta}+\mathcal{H}_{m}(\phi)}{\mathcal{R}(\phi,\theta)},
\label{s-1}\ea
in which
\ba
\mathcal{H}_{m}(\phi)\equiv\frac{\alpha(\phi)^m-\alpha(\phi)^{-m}}{\alpha(\phi)-\alpha(\phi)^{-1}},
\label{calh}\\
\mathcal{R}(\phi,\theta)\equiv\alpha(\phi)^{m+1}+\alpha(\phi)^{-m-1}-2\cos \theta.
\label{calr}\ea
Substituting (\ref{s0}) into (\ref{Guu}) and using (\ref{Gdd}), we find
\ba\fl
{\mathbf{G}}_{\ell,U;\ell',U}=-{\mathbf{G}}_{\ell,D;\ell',D}=-\frac 1 N\sum_{\phi}\,\re^{\ri(\ell-\ell')\phi}2\ri z'\sin\phi \,S_0=\frac 1 N\sum_{\phi}\,\frac{\re^{\ri(\ell-\ell')\phi}\mathcal{D}(\phi)}{z\cal R(\phi,\theta)},
\label{GGuu}\ea
in which
\be
\mathcal{D}(\phi)\equiv-\frac{2\ri z'\sin\phi}{1-z'{}^2}\mathcal{H}_{m+1}(\phi).
\label{cald}\ee
Substituting (\ref{s0}) and (\ref{s-1}) into (\ref{Gud}) and also using (\ref{Gdu}), we obtain
\ba\fl
{\mathbf{G}}_{\ell,U;\ell',D}=-\overline{\mathbf{G}}_{\ell',D;\ell,U}&=&\frac {\re^{\ri\theta}}{N}\sum_{\phi}\re^{\ri(\ell-\ell')\phi} 
[(1-z'{}^2)S_{-1}-z |1+z'\re^{\ri\phi}|^2 S_0]
\nonumber\\
&=&\frac 1 N\sum_{\phi}\,\frac{\re^{\ri(\ell-\ell')\phi}}{\mathcal{R}(\phi,\theta)}\left[z^{-1}-\re^{\ri\theta}\mathcal{A}(\phi)\right],
\label{GGud}\ea
where we have used $\re^{\ri m\psi}=\re^{-\ri\psi}\re^{\ri\theta}$ and
\be
\mathcal{A}(\phi)\equiv\frac{|1+z'\re^{\ri\phi}|^2}{1-z'{}^2}\mathcal{H}_{m+1}(\phi)-
\frac 1 z\,\mathcal{H}_{m}(\phi).
\label{cala}\ee

As we need to calculate the determinant
\be
|\mathbf{m}|=|\bI'+\mathbf{G}\mathbf{y}|,\quad \mathbf{m}\equiv\bI'+\mathbf{G}\mathbf{y},\quad
\ee
in terms of the off-diagonal matrix $\mathbf{y}$ given in (\ref{y}) or (\ref{yy}), we find it convenient to also introduce the function
\be
\mathcal{F}(\phi)\equiv\frac{|1-z'\re^{\ri\phi}|^2}{1-z'{}^2}\mathcal{H}_{m+1}(\phi)-
z\,\mathcal{H}_{m}(\phi).
\label{calf}\ee
It is easy to verify the relations
\ba
\mathcal{A}(\phi)\mathcal{F}(\phi)=1-\mathcal{D}(\phi)^2,
\label{relaAFD}\\
z\mathcal{A}(\phi)+z^{-1}\mathcal{F}(\phi)=\mathcal{R}(\phi,\theta)+2\cos \theta.
\label{relation}\ea
Indeed, noting that
\ba
z\frac{|1+z'\re^{\ri\phi}|^2}{1-z'{}^2}+\frac{1}{z}\frac{|1-z'\re^{\ri\phi}|^2}{1-z'{}^2}
=\alpha(\phi)+\alpha(\phi)^{-1},\nonumber\\
\frac{|1+z'\re^{\ri\phi}|^2}{1-z'{}^2}\frac{|1-z'\re^{\ri\phi}|^2}{1-z'{}^2}+\left[-\frac{2\ri z'\sin\phi}{1-z'{}^2}\right]^2=1,
\ea
the checking of (\ref{relaAFD}) and (\ref{relation}) reduces to verifying simple relations involving only $\alpha(\phi)$.

To shorten the notations, we shall from now on use the abbreviations
\be
\mathcal{R}_k\equiv\mathcal{R}(\phi_k,\theta),\quad
\mathcal{D}_k\equiv\mathcal{D}(\phi_k),\quad
\mathcal{A}_k\equiv\mathcal{A}(\phi_k),\quad
\mathcal{F}_k\equiv\mathcal{F}(\phi_k),
\ee
which are defined in (\ref{calr}), (\ref{cald}), (\ref{cala}) and (\ref{calf})
with $\phi=\phi_k\equiv(2\pi k+\bar\theta)/N$ as in (\ref{Fourier}) and
$\alpha(\phi)=\alpha(\phi_k)$ given in (\ref{alpha}). All these functions are
periodic in $k$ with period $N$. Then
\ba
\fl{\mathbf{m}}_{\ell,D;\ell',D}&=\delta_{\ell,\ell'}+{\mathbf{G}}_{\ell,D;\ell',U}{\mathbf{y}}_{\ell',U;\ell',D}
=\delta_{\ell,\ell'}+\sum_{k=0}^{N-1}\re^{\ri(\ell-\ell')\phi_k}\frac{z\mathcal{A}_k-\re^{\ri\theta}}{N\mathcal{R}_k}
(z^{n-1}\delta_{\ell',N}-1),
\nonumber\\
\fl{\mathbf{m}}_{\ell,D;\ell',U}&={\mathbf{G}}_{\ell,D;\ell',D}{\mathbf{y}}_{\ell',D;\ell',U}
=\sum_{k=0}^{N-1}\re^{\ri(\ell-\ell')\phi_k}\frac{\mathcal{D}_k\re^{-\ri\theta}}{N\mathcal{R}_k}
(z^{n-1}\delta_{\ell',N}-1),
\nonumber\\
\fl{\mathbf{m}}_{\ell,U;\ell',D}&={\mathbf{G}}_{\ell,U;\ell',U}{\mathbf{y}}_{\ell',U;\ell',D}
=\sum_{k=0}^{N-1}\re^{\ri(\ell-\ell')\phi_k}\frac{\mathcal{D}_k\re^{\ri\theta}}{N\mathcal{R}_k}
(z^{n-1}\delta_{\ell',N}-1),
\nonumber\\
\fl{\mathbf{m}}_{\ell,U;\ell',U}&=\delta_{\ell,\ell'}+{\mathbf{G}}_{\ell,U;\ell',D}{\mathbf{y}}_{\ell',D;\ell',U}
=\delta_{\ell,\ell'}+\sum_{k=0}^{N-1}\re^{\ri(\ell-\ell')\phi_k}\frac{z\mathcal{A}_k-\re^{\ri\theta}}{N\mathcal{R}_k}
(z^{n-1}\delta_{\ell',N}-1).
\label{GG}\ea
Using (\ref{relation}), the diagonal elements of $\mathbf{m}$ can be rewritten as
\ba
{\mathbf{m}}_{\ell,D;\ell,D}&=
\sum_{k=0}^{N-1}\frac{(z^{-1}\mathcal{F}_k-\re^{-\ri\theta})+
z^{n-1}\delta_{\ell,N}(z\mathcal{A}_k-\re^{\ri\theta})}{N\mathcal{R}_k},
\nonumber\\
{\mathbf{m}}_{\ell,U;\ell,U}&=
\sum_{k=0}^{N-1}\frac{(z^{-1}\mathcal{F}_k-\re^{\ri\theta})+
z^{n-1}\delta_{\ell,N}(z\mathcal{A}_k-\re^{-\ri\theta})}{N\mathcal{R}_k}.
\label{GGud1}\ea

With the results (\ref{GG}) and (\ref{GGud1}), we can set up the $2N\times2N$ determinant
\be
|\bI'+\mathbf{G}\mathbf{y}|=\left|\begin{array}{cc}
\mathbf{m}_{DD}&\mathbf{m}_{DU}\\
\mathbf{m}_{UD}&\mathbf{m}_{UU}\end{array}\right|.
\label{mat}\ee
For $N=1$ this can be easily evaluated, but already for $N=3$ it becomes a massive computation that can be programmed in Maple, eliminating first the $\mathcal{R}_k$ using (\ref{relation}) and after simplifying the $\mathcal{D}_k^{\;2}$ using (\ref{relaAFD}). Thus one recovers the results from the even more tedious method of the previous section. But, again, there is no need to go into more details, as there is better way that gives the results for general $N$.

Applying the Fourier transform similarity transformation (\ref{Fourier}) to the four
$N\times N$ submatrices in (\ref{mat}), the determinant becomes
\be
|\bI'+\mathbf{G}\mathbf{y}|=\left|\begin{array}{cc}
\hat\mathbf{m}_{DD}&\hat\mathbf{m}_{DU}\\
\hat\mathbf{m}_{UD}&\hat\mathbf{m}_{UU}\end{array}\right|,
\label{matF}\ee
with
\ba\fl
\hat\mathbf{m}_{k,D;k',D}&=\zeta\mathcal{A}_k^{+}+\mathcal{F}_k^{-}\delta_{k,k'},
\quad
\hat\mathbf{m}_{k,D;k',U}&=\zeta\mathcal{D}_k^{-}-\mathcal{D}_k^{-}\delta_{k,k'},
\nonumber\\ \fl
\hat\mathbf{m}_{k,U;k',D}&=\zeta\mathcal{D}_k^{+}-\mathcal{D}_k^{+}\delta_{k,k'},
\quad
\hat\mathbf{m}_{k,U;k',U}&=\zeta\mathcal{A}_k^{-}+\mathcal{F}_k^{+}\delta_{k,k'},
\label{GGG}\ea
where, as easily seen from (\ref{GG}) and (\ref{GGud1})
\be\fl
\mathcal{A}_k^{\pm}=\frac{z\mathcal{A}_k-\re^{\pm\ri\theta}}{\mathcal{R}_k},\quad
\mathcal{F}_k^{\pm}=\frac{z^{-1}\mathcal{F}_k-\re^{\pm\ri\theta}}{\mathcal{R}_k},\quad
\mathcal{D}_k^{\pm}=\frac{\mathcal{D}_k\re^{\pm\ri\theta}}{\mathcal{R}_k},\quad
\zeta\equiv\frac{z^{n-1}}{N}.
\label{AFDpm}\ee

As a representative example, the resulting determinant for $N=3$ is
\be\fl
\left|\begin{array}{cccccc}
\zeta\mathcal{A}_0^{+}+\mathcal{F}_0^{-}&\zeta\mathcal{A}_0^{+}&\zeta\mathcal{A}_0^{+}
&\zeta\mathcal{D}_0^{-}-\mathcal{D}_0^{-}&\zeta\mathcal{D}_0^{-}&\zeta\mathcal{D}_0^{-}\\
\zeta\mathcal{A}_1^{+}&\zeta\mathcal{A}_1^{+}+\mathcal{F}_1^{-}&\zeta\mathcal{A}_1^{+}
&\zeta\mathcal{D}_1^{-}&\zeta\mathcal{D}_1^{-}-\mathcal{D}_1^{-}&\zeta\mathcal{D}_1^{-}\\
\zeta\mathcal{A}_2^{+}&\zeta\mathcal{A}_2^{+}&\zeta\mathcal{A}_2^{+}+\mathcal{F}_2^{-}
&\zeta\mathcal{D}_2^{-}&\zeta\mathcal{D}_2^{-}&\zeta\mathcal{D}_2^{-}-\mathcal{D}_2^{-}\\
\zeta\mathcal{D}_0^{+}-\mathcal{D}_0^{+}&\zeta\mathcal{D}_0^{+}&\zeta\mathcal{D}_0^{+}
&\zeta\mathcal{A}_0^{-}+\mathcal{F}_0^{+}&\zeta\mathcal{A}_0^{-}&\zeta\mathcal{A}_0^{-}\\
\zeta\mathcal{D}_1^{+}&\zeta\mathcal{D}_1^{+}-\mathcal{D}_1^{+}&\zeta\mathcal{D}_1^{+}
&\zeta\mathcal{A}_1^{-}&\zeta\mathcal{A}_1^{-}+\mathcal{F}_1^{+}&\zeta\mathcal{A}_1^{-}\\
\zeta\mathcal{D}_2^{+}&\zeta\mathcal{D}_2^{+}&\zeta\mathcal{D}_2^{+}-\mathcal{D}_2^{+}
&\zeta\mathcal{A}_2^{-}&\zeta\mathcal{A}_2^{-}&\zeta\mathcal{A}_2^{-}+\mathcal{F}_2^{+}
\end{array}\right|.
\label{detN3}\ee
We can easily verify that the determinant $|\bI'+\mathbf{G}\mathbf{y}|$ is a polynomial
in $\zeta$ of degree at most 2. Leaving columns 1 and $N+1$ unchanged, ($N=3$ in the
example), we subtract the first column from columns 2 to $N$ and subtract column $N+1$
from columns $N+2$ to $2N$. After that only columns 1 and $N+1$ contain $\zeta$,
showing that there cannot be third powers of $\zeta$ and higher.

Next we choose $0D,0U,1D,1U,\ldots,(N\!-\!1)D,(N\!-\!1)U$ for the ordering of the matrix indices
of $\hat\mathbf{m}$, so that example (\ref{detN3}) with $N=3$ becomes
\be\fl
\left|\begin{array}{cccccc}
\zeta\mathcal{A}_0^{+}+\mathcal{F}_0^{-}&\zeta\mathcal{D}_0^{-}-\mathcal{D}_0^{-}
&\zeta\mathcal{A}_0^{+}&\zeta\mathcal{D}_0^{-}
&\zeta\mathcal{A}_0^{+}&\zeta\mathcal{D}_0^{-}\\
\zeta\mathcal{D}_0^{+}-\mathcal{D}_0^{+}&\zeta\mathcal{A}_0^{-}+\mathcal{F}_0^{+}
&\zeta\mathcal{D}_0^{+}&\zeta\mathcal{A}_0^{-}
&\zeta\mathcal{D}_0^{+}&\zeta\mathcal{A}_0^{-}\\
\zeta\mathcal{A}_1^{+}&\zeta\mathcal{D}_1^{-}
&\zeta\mathcal{A}_1^{+}+\mathcal{F}_1^{-}&\zeta\mathcal{D}_1^{-}-\mathcal{D}_1^{-}
&\zeta\mathcal{A}_1^{+}&\zeta\mathcal{D}_1^{-}\\
\zeta\mathcal{D}_1^{+}&\zeta\mathcal{A}_1^{-}
&\zeta\mathcal{D}_1^{+}-\mathcal{D}_1^{+}&\zeta\mathcal{A}_1^{-}+\mathcal{F}_1^{+}
&\zeta\mathcal{D}_1^{+}&\zeta\mathcal{A}_1^{-}\\
\zeta\mathcal{A}_2^{+}&\zeta\mathcal{D}_2^{-}
&\zeta\mathcal{A}_2^{+}&\zeta\mathcal{D}_2^{-}
&\zeta\mathcal{A}_2^{+}+\mathcal{F}_2^{-}&\zeta\mathcal{D}_2^{-}-\mathcal{D}_2^{-}\\
\zeta\mathcal{D}_2^{+}&\zeta\mathcal{A}_2^{-}
&\zeta\mathcal{D}_2^{+}&\zeta\mathcal{A}_2^{-}
&\zeta\mathcal{D}_2^{+}-\mathcal{D}_2^{+}&\zeta\mathcal{A}_2^{-}+\mathcal{F}_2^{+}
\end{array}\right|.
\label{detN3a}\ee
Then, after setting $\zeta=0$, the determinant simply reduces to a product of determinants
of $2\times2$ matrices, $\prod_{k=0}^{N-1} d_k^{(0)}$ with
\be
d_k^{(0)}\equiv
\left|\begin{array}{cc}
\mathcal{F}_k^{-}&-\mathcal{D}_k^{-}\\
-\mathcal{D}_k^{+}&\mathcal{F}_k^{+}
\end{array}\right|
=\frac{|z^{-1}\mathcal{F}_k-\re^{\ri\theta}|^2-\mathcal{D}_k^{\;2}}{\mathcal{R}_k^{\;2}}
=\frac{\mathcal{F}_k}{z\mathcal{R}_k},
\label{d0k}\ee
upon using (\ref{relaAFD}) and (\ref{relation}) in the last step.

The term linear in $\zeta$ is the sum of all $4N^2$ determinants with all but one of the $\zeta$'s
set to zero. The one $\zeta$ left can be in any entry, but, if we don't select that $\zeta$ within
one of the $N$ $2\times2$ block matrices along the diagonal, its matrix element has indices
corresponding to two different blocks, say $k$ and $l$. Then we must consider the
$4\times4$ determinant
\be
M_{kl}=
\left|\begin{array}{cccc}
\mathcal{F}_k^{-}&-\mathcal{D}_k^{-}&\zeta\mathcal{A}_k^{+}&\zeta\mathcal{D}_k^{-}\\
-\mathcal{D}_k^{+}&\mathcal{F}_k^{+}&\zeta\mathcal{D}_k^{+}&\zeta\mathcal{A}_k^{-}\\
\zeta\mathcal{A}_l^{+}&\zeta\mathcal{D}_l^{-}&\mathcal{F}_l^{-}&-\mathcal{D}_l^{-}\\
\zeta\mathcal{D}_l^{+}&\zeta\mathcal{A}_l^{-}&\-\mathcal{D}_l^{+}&\mathcal{F}_l^{+}
\end{array}\right|,
\label{mkl}\ee
with all $\zeta$'s but one set zero. There are eight choices of keeping one $\zeta$ in $M_{kl}$.
It is easily seen that, for any of these choices, the determinant of the $3\times3$ minor of the
chosen $\zeta$ entry vanishes, and so does the corresponding $4\times4$ determinant.
We are left with the $4N$ choices of $\zeta$ within a diagonal $2\times2$ block. Hence,
we get a sum of $\prod_k d_k^{(0)}$ with one factor $d_k^{(0)}$ replaced by $\zeta d_k^{(1)}$,
where, using (\ref{relaAFD}) and (\ref{relation}),
\ba\fl
d_k^{(1)}&\equiv
\left|\begin{array}{cc}
\mathcal{A}_k^{+}&0\\
0&\mathcal{F}_k^{+}
\end{array}\right|+
\left|\begin{array}{cc}
0&\mathcal{D}_k^{-}\\
-\mathcal{D}_k^{+}&0
\end{array}\right|+
\left|\begin{array}{cc}
0&-\mathcal{D}_k^{-}\\
\mathcal{D}_k^{+}&0
\end{array}\right|+
\left|\begin{array}{cc}
\mathcal{F}_k^{-}&0\\
0&\mathcal{A}_k^{-}
\end{array}\right|
\nonumber\\
\fl&=\frac{2\mathcal{A}_k\mathcal{F}_k-2(z\mathcal{A}_k+z^{-1}\mathcal{F}_k)\cos(\theta)+2\cos(2\theta)
+2\mathcal{D}_k^{\;2}}{\mathcal{R}_k^{\;2}}
=-\frac{2\cos\theta}{\mathcal{R}_k}.
\label{d1k}\ea

The term quadratic in $\zeta$ has three contributions.\footnote{The reader can easily verify
that the other types, with only one $\zeta$ outside the $2\times2$ blocks or with both
$\zeta$'s outside with indices corresponding to three or four blocks, lead to singular $2\times2$
or $4\times4$ minors, for example by striking out the two rows and two columns
of the two chosen $\zeta$ entries in (\ref{detN3a}).} The first has both $\zeta$'s
in the same $2\times2$ matrix, so one gets the sum of $\prod_k d_k^{(0)}$ with
one factor replaced by $\zeta^2 d_k^{(2,1)}$ with
\be\fl
d_k^{(2,1)}\equiv
\left|\begin{array}{cc}
\mathcal{A}_k^{+}&0\\
0&\mathcal{A}_k^{-}
\end{array}\right|+
\left|\begin{array}{cc}
0&\mathcal{D}_k^{-}\\
\mathcal{D}_k^{+}&0
\end{array}\right|
=\frac{z^2\mathcal{A}_k^{\;2}-2z\mathcal{A}_k\cos\theta+1-\mathcal{D}_k^{\;2}}
{\mathcal{R}_k^{\;2}}=\frac{z\mathcal{A}_k}{\mathcal{R}_k}.
\label{d21k}\ee
The second contribution has one $\zeta$ each in two different $2\times2$ matrices. Then
we must replace $d_k^{(0)}d_l^{(0)}$ in  $\prod_k d_k^{(0)}$ by $\zeta^2 d_{kl}^{(2,2)}$
for all possible pairs $(k,l)$, where
\be
d_{kl}^{(2,2)}=d_k^{(1)}d_l^{(1)}=\frac{4\cos^2\theta}{\mathcal{R}_k\mathcal{R}_l}.
\label{d22k}\ee
The third contribution results from two $\zeta$'s in entries outside the $2\times2$ blocks at
$\zeta=0$, but with matrix indices belonging to the same two such blocks, say the ones
with $k$ and $l$.
Now we must replace $d_k^{(0)}d_l^{(0)}$ by $\zeta^2 d_{kl}^{(2,3)}$, with
$d_{kl}^{(2,3)}$ the coefficient of $\zeta^2$ in the expansion of (\ref{mkl}),
with one $\zeta$ in the upper-right quarter of (\ref{mkl}) and the other in the lower left quarter.
Collecting the 16 non-zero contributions, substituting (\ref{AFDpm}), followed by (\ref{relaAFD})
and finally (\ref{relation}), we find
\ba\fl
d_{kl}^{(2,3)}=&-(\mathcal{A}_k^{+}\mathcal{F}_k^{+}\mathcal{A}_l^{+}\mathcal{F}_l^{+}
  +\mathcal{A}_k^{-}\mathcal{F}_k^{-}\mathcal{A}_l^{-}\mathcal{F}_l^{-})
-\mathcal{D}_k^{+}\mathcal{D}_k^{-}
  (\mathcal{A}_l^{+}\mathcal{F}_l^{+}+\mathcal{A}_l^{-}\mathcal{F}_l^{-})
\nonumber\\
\fl&-\mathcal{D}_l^{+}\mathcal{D}_l^{-}
  (\mathcal{A}_k^{+}\mathcal{F}_k^{+}+\mathcal{A}_k^{-}\mathcal{F}_k^{-})
-2\mathcal{D}_k^{+}\mathcal{D}_k^{-}\mathcal{D}_l^{+}\mathcal{D}_l^{-}
\nonumber\\
\fl&-\mathcal{D}_k^{+}\mathcal{D}_l^{-}(\mathcal{A}_k^{+}+\mathcal{F}_k^{-})
 (\mathcal{A}_l^{-}+\mathcal{F}_l^{+})
-\mathcal{D}_k^{-}\mathcal{D}_l^{+}(\mathcal{A}_k^{-}+\mathcal{F}_k^{+})
 (\mathcal{A}_l^{+}+\mathcal{F}_l^{-})
 \nonumber\\
\fl\phantom{d_{kl}^{(2,3)}}=&\frac{2(1-\mathcal{D}_k\mathcal{D}_l-2\cos^2\theta)
(z\mathcal{A}_k+z^{-1}\mathcal{F}_k-2\cos\theta)
(z\mathcal{A}_l+z^{-1}\mathcal{F}_l-2\cos\theta)}
{\mathcal{R}_k^{\;2}\mathcal{R}_l^{\;2}}
 \nonumber\\
\fl\phantom{d_{kl}^{(2,3)}}=&\frac{2(1-\mathcal{D}_k\mathcal{D}_l)-4\cos^2\theta}
{\mathcal{R}_k\mathcal{R}_l}=\frac{2(1-\mathcal{D}_k\mathcal{D}_l)}
{\mathcal{R}_k\mathcal{R}_l}-d_{kl}^{(2,2)}.
\label{d23k}\ea

Collecting the results from (\ref{d0k}), (\ref{d1k}), (\ref{d21k}), (\ref{d22k}), (\ref{d23k}), and
using (\ref{AFDpm}) to eliminate $\zeta$ and then (\ref{relaAFD}) to eliminate
the $\mathcal{A}_k$'s, we have
\ba\fl
|\bI'+\mathbf{G}\mathbf{y}|&=\prod_{k=0}^{N-1}\frac{\mathcal{F}_k}{z\mathcal{R}_k}
\left[1-\sum_{k=0}^{N-1}\frac{2Nz^n\cos\theta-z^{2n}\mathcal{A}_k}{N^2\mathcal{F}_k}
+\sum_{k=0}^{N-2}\sum_{l=k+1}^{N-1}\frac{2z^{2n}(1-\mathcal{D}_k\mathcal{D}_l)}
{N^2\mathcal{F}_k\mathcal{F}_l}\right],
\nonumber\\
\fl&=\prod_{k=0}^{N-1}\frac{\mathcal{F}_k}{z\mathcal{R}_k}
\left[1-\frac{2z^n\cos\theta}{N}\sum_{k=0}^{N-1}\frac{1}{\mathcal{F}_k}
+\frac{z^{2n}}{N^2}\sum_{k=0}^{N-1}\sum_{l=0}^{N-1}\frac{1-\mathcal{D}_k\mathcal{D}_l}
{\mathcal{F}_k\mathcal{F}_l}\right].
\label{DetN}\ea
This we can rewrite as
\be
|\bI'+\mathbf{G}\mathbf{y}|=\frac {z^{n-N}}{N\mathcal{R}_0\mathcal{R}_1\cdots\mathcal{R}_{N-1}}[{\tau}_N(\bar\theta)-{\rho}_N(\bar\theta)\cos\theta ],
\label{GyN}\ee
with
\be
{\tau}_N(\bar\theta)=Nz^{-n}\prod_{k=0}^{N-1}{\mathcal{F}_k}+
\frac{z^n}{N}\prod_{k=0}^{N-1}{\mathcal{F}_k}\left[
\left(\sum_{k=0}^{N-1}\frac{1}{\mathcal{F}_k}\right)^2-
\left(\sum_{k=0}^{N-1}\frac{\mathcal{D}_k}{\mathcal{F}_k}\right)^2\right],
\label{tau23}\ee
and
\be
\rho_N(\bar\theta)=2\prod_{k=0}^{N-1}{\mathcal{F}_k}\sum_{k=0}^{N-1}\frac{1}{\mathcal{F}_k}.
\label{rho23}\ee

Using (\ref{integral}), we find from (\ref{calr}) that
\ba
\int_0^{2\pi}\!\!\rd\theta \,\ln\mathcal{R}(\phi, \theta)=2\pi\ln\alpha(\phi)^{m+1},
\label{intR}\ea
and from (\ref{GyN}), 
\be\fl
\int_0^{2\pi}\!\!\rd\theta \,\ln[{\tau}_N(\bar\theta)-{\rho}_N(\bar\theta)\cos\theta]
=2\pi\ln\Big[\half\Big(\tau_N(\bar\theta)+\sqrt{\tau^2_N(\bar\theta)-\rho^2_N(\bar\theta)}\Big)\Big].
\label{intyG}\ee
From (\ref{dU}), we find using (\ref{intU0}), (\ref{GyN}), (\ref{intR}) and (\ref{intyG}), the result
\ba
\int_0^{2\pi}\!\!\rd\bar\theta\int_0^{2\pi}\!\!\rd\theta|\bU(\theta,{\bar \theta})|=\int_0^{2\pi}\!\!\rd\theta\ln|\bU_0(\theta,{\bar \theta})|+\int_0^{2\pi}\!\!\rd\theta\ln|\bI'+\mathbf{G}\mathbf{y}|
\nonumber\\
\hspace{30pt}=(2\pi)^2[(m+1)N\ln(z'(1-z^2))+\ln (z^{n-N}/N)]
\nonumber\\
\hspace{40pt}+2\pi\int_0^{2\pi}\!\!\rd\bar\theta\ln\Big[\half\Big(\tau_N(\bar\theta)+\sqrt{\tau^2_N(\bar\theta)-\rho^2_N(\bar\theta)}\Big)\Big].
\label{intU}\ea
Now combining (\ref{freeEnergy}) with (\ref{intU}), substituting $z$ and $z'$ from (\ref{tbJ})
and (\ref{B}), we obtain
\ba\fl
-\beta f
={\hat f}_0(N,m,n) +\frac1{4\pi N(m+n)}
\int_0^{2\pi}\!\!\rd{\bar\theta}\ln\Big[\half\Big(\tau_N(\bar\theta)+\sqrt{\tau^2_N(\bar\theta)-\rho^2_N(\bar\theta)}\Big)\Big],
\label{freeEnergy2}\ea
where
\ba\fl
{\hat f}_0(N,m,n)=\frac1{2N(m+n)}\big[N(m+1)\ln\sinh(2\beta J')
+(n-N)\ln\sinh(2\beta J)
\nonumber\\
\hspace{60pt}+(n-2+Nm+2N)\ln 2-\ln N\big].
\label{hf0}\ea

The specific heat is the second derivative of the free energy which is specifically given by (3.35) together with (3.29) on p.~92 in the book \cite{MWbk}. Thus the specific heats shown in paper I
\cite{HJPih}, are the results of differentiating the free energy given by (\ref{freeEnergy2}).
\section{Critical Temperature}
The critical temperature is determined from the equation
\be
\tau_N(0)-\rho_N(0)=0.
\label{Tc0}\ee
For $\bar\theta=0$, (\ref{psi}) becomes $\phi_j=2\pi j/N$, so that from  (\ref{alpha}) we find
$\mathcal{H}(\phi_{N-k})=\mathcal{H}(\phi_{k})$, with the result the same for both solutions
of the quadratic equation (\ref{alpha}) for the $\alpha(\phi)$. Therefore, from (\ref{calf})
and (\ref{cald}) we have
\be
\mathcal{F}_{N-k}=\mathcal{F}_k,\quad
\mathcal{D}_{N-k}=-\mathcal{D}_k,\quad\mathcal{D}_0=0=\mathcal{D}_{N/2},
\label{FDNk}\ee
(with the last equality needed when $N$ is even), implying
\be
\sum_{k=0}^{N-1}\frac{\mathcal{D}_k}{\mathcal{F}_k}=0,\quad\mbox{for }\bar\theta=0,
\ee
so that
\be
\tau_N(0)\pm\rho_N(0)=
\left(Nz^{-n}\prod_{k=0}^{N-1}{\mathcal{F}_k}\right)
\left(1\pm\frac{z^n}{N}\sum_{k=0}^{N-1}\frac{1}{\mathcal{F}_k}\right)^2,
\ee
Consequently, the critical temperature is determined by the equation
\be
\frac{z^n}{N}\sum_{k=0}^{N-1}\frac{1}{\mathcal{F}_k}=1.
\label{Tc}\ee
Now $\bar\theta=0$, we have from (\ref{alpha})
\ba\fl
\frac{\alpha_k^{\vphantom{-1}}+\alpha_k^{-1}}{2}&=r_k\equiv
\frac{(1+z^2)(1+z'{}^2)}{2z(1-z'{}^2)}-\frac{z'(1-z^2)}{z(1-z'{}^2)}\cos\phi_k
\nonumber\\ \fl
&=\frac{\cosh(2\beta J)\cosh(2\beta J')}{\sinh(2\beta J)}
-\frac{\sinh(2\beta J')}{\sinh(2\beta J)}\cos\frac{2\pi k}{N}
=\cosh(\gamma_{2k}),
\label{rk}\ea
where $\gamma_{2k}$ has been introduced before in eq.~(89a) of \cite{Onsager} and
eq.~(51) of \cite{Kaufman}, replacing $n$ there by $N$, so that
\be
\alpha_k=\re^{\gamma_{2k}}=r_k+\sqrt{r_k^2-1}.
\label{alphak}\ee
In two cases the square root disappears, namely for $k=0$, $\phi_0=0$ and
$k=N/2$, $\phi_{N/2}=\pi$ (when $N$ is even). For these cases we find
\ba
\alpha_0=\frac{1-z'}{z(1+z')},\quad \mathcal{F}_0=z\alpha_0^{m+1},
\nonumber\\
\alpha_{N/2}=\frac{1+z'}{z(1-z')},\quad \mathcal{F}_{N/2}=z\alpha_{N/2}^{m+1}.
\label{alpha0pi}\ea
\subsection{The case $N=1$}
When $N=1$, we find from (\ref{Tc}) and (\ref{alpha0pi}), that $T_c$ is given by
\be
\left[\frac{(1-z')}{(1+z')}\right]^{m+1}=z^{n+m}.
\label{Tc1}\ee
This is identical to (2) of paper I \cite{HJPih} for $J'=J$.

\subsection{The case $N=2$}
For $N=2$, again using (\ref{Tc}) and (\ref{alpha0pi}), we find that $T_c$ is given by\ba
\frac{z^{n+m}}{2}\left[\bigg(\frac{1-z'}{1+z'}\bigg)^{m+1}+\bigg(\frac{1+z'}{1-z'}\bigg)^{m+1}\right]
\nonumber\\
=\tanh(\beta J)^{n+m}\cosh[2(m+1)\beta J']=1,
\label{Tc2f}\ea
which is identical to (4) of paper I \cite{HJPih} for $J'=J$.

\subsection{The case $N=3$}
For $N=3$, with $\phi_1=2\pi/3$ and $\phi_2=4\pi/3$, or $\cos\phi_1=\cos\phi_2=-1/2$, we find from (\ref{alphak}) and (\ref{calf}) that
\ba 
\alpha_1=\alpha_2=r+\sqrt{r^2-1},\quad r=\frac{(1+z^2)(1+z'{}^2)+z'(1-z^2)}{2z(1-z'{}^2)}.
\label{alpha2}\\
F_1=F_2=\frac{1+z'+z'{}^2}{1-z'{}^2}
\bigg[\frac{\alpha_1^{m+1}-\alpha_1^{-m-1}}{\alpha_1-\alpha_1^{-1}}\bigg]
-z\bigg[\frac{\alpha_1^{m}-\alpha_1^{-m}}{\alpha_1-\alpha_1^{-1}}\bigg],
\label{calfn}\ea
Thus we find upon using (\ref{Tc}) and (\ref{alpha0pi}) that the condition for $T_c$ becomes
\be
F_1\bigg[1-\frac{z^{m+n}}3\Big(\frac{1+z'}{1-z'}\Big)^{m+1}\bigg]=\frac{2 z^n}{3}.
\label{Tc3f}\ee
For $J'=J$,  (\ref{Tc3f}), (\ref{calfn}), and (\ref{alpha2}) are the same as (5), (6) and (7) in paper I \cite{HJPih}.
\subsection{The general case}
Now we can again use (\ref{FDNk}), (\ref{Tc}) and (\ref{alpha0pi}). For $N=2p+1$ odd we find
\be
\sum_{k=1}^{p}\frac{1}{\mathcal{F}_k}=\frac{N}{2z^n}
-\frac{z^m}{2}\bigg(\frac{1+z'}{1-z'}\bigg)^{m+1},
\ee
whereas for $N=2p$ even,
\be
\sum_{k=1}^{p-1}\frac{1}{\mathcal{F}_k}=\frac{N}{2z^n}
-\frac{z^m}{2}\bigg(\frac{1+z'}{1-z'}\bigg)^{m+1}
-\frac{z^m}{2}\bigg(\frac{1-z'}{1+z'}\bigg)^{m+1}.
\ee
Here we have to use (\ref{calf}) with (\ref{alphak}) and (\ref{rk}) to calculate the $\mathcal{F}_k$.
It is not necessary to calculate $\alpha_k$ for that, as  $\mathcal{F}_k$ is given in terms of
 $\mathcal{H}_m(\phi_k)=U_{m-1}(r_k)$, a Chebyshev polynomial of the second kind in $r_k$
 defined in (\ref{rk}). This is also true for the $\mathcal{F}_k$ and $\mathcal{D}_k$ with
 $\bar\theta\ne0$ in (\ref{tau23}) and (\ref{rho23}) needed to calculate (\ref{freeEnergy2}).

\end{document}